\documentclass{aastex}
\usepackage{spr-astr-addons}

\usepackage{url}
\RequirePackage{color}
\usepackage{longtable}
\usepackage{natbib}
\usepackage{amsmath}
\usepackage{txfonts}
\usepackage{graphicx}
\usepackage{amssymb}
\usepackage[labelfont=bf]{caption}
\begin{document}

\title{Tidal Dwarf Galaxies at Intermediate Redshifts}
\slugcomment{Submitted to Ap\&SS}
%% Running heads
\shorttitle{Tidal dwarf galaxies at intermediate redshifts}
\shortauthors{Wen et al.}

%\author{Zhang-Zheng Wen\altaffilmark{1,2}$^\dagger$, Xian-Zhong Zheng\altaffilmark{1}, Ying-He Zhao\altaffilmark{1}, Yu Gao\altaffilmark{1}} 

%\altaffiltext{1}{Purple Mountain Observatory, Chinese Academy of Sciences, Nanjing 210008, China}
%\altaffiltext{2}{Graduate University of Chinese Academy of Sciences, Beijing 100049, China}
%\altaffiltext{$^\dagger$}{zzwen@pmo.ac.cn}

\author{Zhang-Zheng Wen\altaffilmark{1}$^\dagger$}
%\affil{Purple Mountain Observatory, Chinese Academy of Sciences, Nanjing 210008, China}
\and 
\author{Xian-Zhong Zheng\altaffilmark{}}
%\affil{Purple Mountain Observatory, Chinese Academy of Sciences, Nanjing 210008, China}
\and
\author{Ying-He Zhao\altaffilmark{}}
%\affil{Purple Mountain Observatory, Chinese Academy of Sciences, Nanjing 210008, China}
\and
\author{Yu Gao\altaffilmark{}}
\affil{Purple Mountain Observatory, Chinese Academy of Sciences, Nanjing 210008, China}
%\email{asdf@asdf}

\altaffiltext{1}{Graduate University of Chinese Academy of Sciences, Beijing 100049, China}

\altaffiltext{$^\dagger$}{e-mail: zzwen@pmo.ac.cn}

\begin{abstract}
 We present the first attempt at measuring the production rate of tidal dwarf galaxies (TDGs) and estimating their contribution to the overall dwarf population. Using HST/ACS deep imaging data from GOODS and GEMS surveys in conjunction with photometric redshifts from COMBO-17 survey, we performed a  morphological analysis for a sample of merging/interacting galaxies in the Extended Chandra Deep Field South and identified tidal dwarf candidates in the rest-frame optical bands. We estimated a production rate about $1.4\times 10^{-5}$ per Gyr per comoving volume for long-lived TDGs with stellar mass $3\times 10^{8-9}$\,M$_\odot$ at $0.5<z<1.1$.  Together with galaxy merger rates and %a survival rate of 25\% for TDGs 
TDG survival rate from the literature, our results suggest that only a marginal fraction (less than 10\%) of dwarf galaxies in the local universe could be tidally-originated. TDGs in our sample are on average bluer than their host galaxies in the optical. Stellar population modelling of optical to near-infrared spectral energy distributions (SEDs) for two TDGs favors a burst component with age 400/200\,Myr and stellar mass 40\%/26\% of the total, indicating that a young stellar population newly formed in TDGs. This is consistent with the episodic star formation histories found for nearby TDGs.
\end{abstract}


%\section*{}f
%\label{sec:intro}

\section{Introduction}

  The last two decades have witnessed much progress in understanding galaxy formation and evolution on both observational and theoretical sides. While the formation and evolution of massive galaxies have been empirically explored in great detail out to $z\sim 4$ with cosmic deep surveys, a complete picture for the origin of dwarf galaxies is still missing. 
Generally speaking, a dwarf galaxy can be formed from either collapse of primordial gas cloud in the framework of cosmology (i.e., classical dwarf), or materials driven by tidal force away from massive galaxies in interactions and/or mergers \citep[so-called ``tidal dwarf''][]{Zwicky56}. 
The classical dwarf galaxies are characterized by small size and dominated by dark matter halo \citep{Aaronson83,Mateo93,Simon07,Geha09}. They are metal poor because of inefficient chemical enrichment in shallow gravitational potential which keeps little metal against supernova winds \citep{Tremonti04}, and thus sensitive for testing physical mechanisms driving galaxy evolution (e.g. supernova feedback).
% As analogs of first galaxies, the classical dwarf galaxies can be resolved star by star in the Local Group \citep{Weisz08} and provide insight into the processes in the early universe.
Moreover, the cosmologically-originated dwarf galaxies give rise to a well-known challenge to the theory of galaxy formation, i.e., the ``missing satellites problem'' \citep{Klypin99, Moore99}. 
In contrary, tidal dwarf galaxies (TDGs) are believed to contain likely no dark matter halo (Barnes \& Hernquist~1992; Braine et al.~2001; Wetzstein et al.~2007)
%\citep{Barnes92,Braine01,Wetzstein07} \
and be metal rich \citep{Duc00}. They often have episodic star formation histories (SFHs; Weisz et al.~2008; Sabbi et al.~2008), compared to the constant SFH for classical dwarfs (Marconi et al.~1995; Tolstoy et al.~2009).
Understanding of the nature of TDGs and their contribution to the local dwarf population is therefore an important issue in dwarf galaxy astrophysics.

 Observations and numerical simulations manifest that TDGs are formed from recycled materials previously expelled from massive galaxies in tidal interaction (Duc et al.~2000; Weilbacher et al.~2000; Bournaud \& Duc~2006; Wetzstein et al.~2007).
%{\it Small scale gravitational instability drives tidal tail to form stellar clumps in which gas falls in \citep{Barnes92}; or gravitationally-unstable gas gets cooled and condensed at first (Elmegreen et al.~1993; Wetzstein et al.~2007). TDGs usually have a mass of several 10$^7$\,M$_\odot$ to 10$^8$\,M$_\odot$, even up to $10^9$\,M$_\odot$ \citep[e.g. A245N,][]{Duc00} at the tip of tidal tails where a large amount of gas is accumulated (Bournaud et al.~2003).} 
Small TDGs of mass $\sim 10^7$\,M$_\odot$ may be formed from tidal tails (gas and stars) caused by small scale gravitational instability \citep{Barnes92,Wetzstein07}.  Massive TDGs with mass $\sim 10^8 - 10^9$\,M$_\odot$ are usually made up at the tip of a tidal tail by accumulating materials originally from outer disks of the progenitor spiral galaxies \citep{Elmegreen93,Duc04}. 
%These materials will collapse without fragmenting into many unbound smaller objects because of inner high velocity dispersion.} 
According to the hierarchical scenario, massive galaxies were gradually assembled through a number of merger events (Eisenstein et al.~2005, and references therein), providing potential space for numerous TDGs produced over cosmic time.
Indeed, galaxy merger rate increases rapidly with redshift (e.g. Abraham et al.~1996; Neuschaefer et al.~1997; Conselice et al.~2003; Lavery et al.~2004, Straughn et al.~2006; Lotz et al.~2006). A typical $L^\ast$ galaxy in the present day is expected to has undergone $\sim 4$ major mergers since $z\sim 3$ \citep{Conselice06}. From this point of view, a substantial fraction of local dwarf galaxies might be tidally originated. However, TDGs are exclusively associated with tidal tails driven by rotation-support systems. They are also affected by the tidal friction with their parent galaxies. It is likely that only a small fraction of TDGs are able to live longer than 10\,Gyr, depending on their mass and distance to the parent galaxies. These uncertainties leave little room for a meaningful estimate of production rate for TDGs from theories or numerical simulations \citep{Okazaki00, Bournaud06}. Observational examination is therefore essential to determining TDG production rates \citep[e.g. in compact environment][]{Hunsberger96}.

 While a complete and systematic investigation of local TDGs is not available yet, the current generation cosmic deep surveys provide capability for a census of TDGs in distant universe. Imaging with Hubble Space Telescope (HST) provides spatial resolution and sensitivity to determine morphologies of distant galaxies and detect faint signs of merging/interacting. We combine HST Advanced Camera for Survey (ACS) wide-area deep imaging data from the Great Observatories Origins Deep Survey south (GOODS-S) and the Galaxy Evolution from Morphology and SED (GEMS) survey with redshift catalog from the COMBO-17 survey to examine merging/interacting galaxies at intermediate redshifts ($0.5<z<1.1$) in the Extended Chandra Deep Field South (ECDFS) and estimate TDG production rates in the examined cosmic epoch. 

  This paper is organized as follows: in Section 2 and 3 we describe the selection for the tidal dwarf galaxy sample and the results of our analysis, including the estimate of production rate of TDGs. Section 4 presents stellar population analysis of broad-band SEDs for two TDGs in the GOODS-S and finally in Section 5 we present our discussion and conclusions. Throughout we adopt the standard concordance cosmology with $\Omega_{\rm m} = 0.3$, $\Omega_\Lambda=0.7$ and use $H_0 = 70$\,km\,s$^{-1}$\,Mpc$^{-1}$. All magnitudes are given in the AB system.

\section{Sample and Data}

\subsection{HST/ACS Data}

  GOODS-S and GEMS HST/ACS imaging data are publicly available. The GOODS-S observations consist of $F435W(B)$, $F606W(V)$, $F775W(i)$ and $F850LP(z)$ band imaging over the central 160 square arcminute area of the ECDFS. The GEMS observations contain $F606W$ and $F850LP$ imaging of the outer $28\arcmin \times 28\arcmin$ region of the ECDFS. We use the released mosaic images and photometric catalogs from the GOODS-S \citep[version 2.0,][]{Giavalisco04} and GEMS (Rix et al. 2004) surveys for our analysis. All images have a pixel size of $0\farcs 03$ and a spatial resolution of $0\farcs 12$.  The GOODS-S $B$- $V$-, $i$- and $z$-band observations reach a 5\,$\sigma$ point-source depth of 28.6, 28.6, 27.9 and 27.4\,mag respectively, compared to 28.3 and 27.1\,mag for GEMS $V-$ and $z$-band observations. Details of observations and data reduction can be found in corresponding references \citep{Giavalisco04, Rix04}.

We adopt photometric redshifts from the COMBO-17 survey \citep{Wolf03} for more than 8000 galaxies with $R<24.1$ in GOODS-S and GEMS sky coverage. Our morphological analysis is based on the $z$-band imaging data which allows us to investigate morphologies in the rest-frame optical for galaxies out to z=1.1.

\subsection{Sample of Merging/Interacting Galaxies}

In practice a TDG candidate can be securely recognized if it is tied with a massive galaxy via a tidal tail. Otherwise it would be difficult to distinguish an isolated TDG from a cosmologically-originated satellite galaxy. In the merging/interacting process, a tidal tail eventually fades away several 10$^8$\,yrs after first perigalacticon while gas and stars on the tail disperse into the intergalactic medium or fall back onto its parent galaxy \citep{Mihos95}. Brighter tidal tails usually appear in early stage of the Toomre sequence \citep{Toomre72}. We thus select only the merging/interacting galaxies with apparent tidal tails to make a parent sample for TDG identification.  Moreover, only bright tidal tails and massive TDGs are detectable at such great cosmic distance due to the cosmic dimming effect.

\begin{figure}[ht]
  \centering
 \includegraphics[width=0.48\textwidth,bb=85 345 550 700]{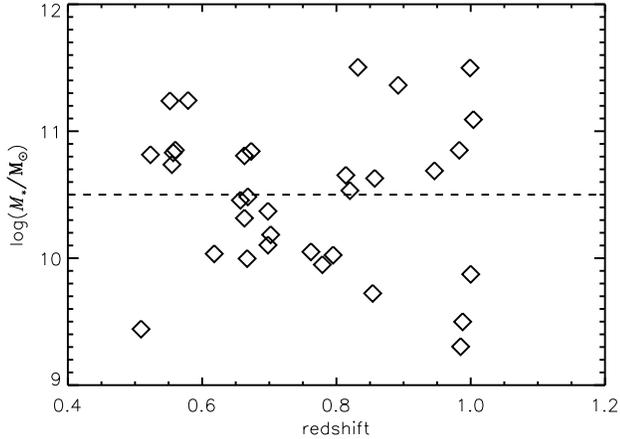}
 \caption{The distribution of stellar mass versus redshift for our sample of 33 merging/interacting galaxies at $0.5<z<1.1$. The stellar mass is derived from luminosity in combination with color following Bell et al. [38]. The dashed line represents the mass cut at $3\times10^{10}$\,M$_\odot$.\label{fig1}}
\end{figure}

A visual selection of merging/interacting galaxies in the ECDFS has been carried out by Elmegreen et al. (2007) using the same HST imaging data as we use here. 
Of 8565 galaxies with $z<1.4$, about 300 show interaction features such as tails, bridges, tidal arm, diffuse plumes and close pairs. There are 33 of the 300 galaxies in the redshift range $0.5<z<1.1$ with clear tidal tails of length $>0.'6$. Such strong tidal features appear exclusively in major mergers. We took the 33 galaxies with strong tidal tails at $0.5<z<1.1$ as our sample for further TDG identification. The redshift distribution of the sample is shown in Figure~\ref{fig1}, with a median redshift of 0.7.

\section{Analysis and Results}

\subsection{Stellar Mass and Surface Brightness}

 We estimated stellar mass using the color-luminosity method \citep{Bell01, Bell03}. To derive the rest-frame color and luminosity, a set of galaxy spectral templates from Bruzual \& Charlot (2003, BC03) are used to match the observed $V-z$ color of every target. The template set is composed of spectra of a galaxy with solar metallicity and e-folding ($\tau$=1\,Gyr) SFH aging from 10\,Myr to 20\,Gyr. The best-fit template is selected using the linear least-squares fitting method. $K$-corrections are derived from the best-fit template to convert the observed $V$ and $z$-band luminosities (corresponding to rest-frame $U$ and $V$-band for a $z=0.7$ galaxy) into rest-frame Johnson $B$ and $V$-band luminosities, respectively. This minimizes the uncertainties of $K$-corrections for galaxies at $z\sim 0.7$ \citep[see][for more details]{Hammer01}. The relation between mass-to-light ratio $M/L$ and $B-V$ color of the integrated stellar populations given in Bell et al. (2003) is adopted to derive stellar mass from $V$-band luminosity for our sample galaxies. A Charbier initial mass function (IMF) \citep{Chabrier03} with mass range from $0.1$\,M$_\odot$ to $100$\,M$_\odot$ is assumed and the calculation of stellar mass has been corrected according to Bernardi et al. (2010). The estimated stellar mass has a typical error of 0.1-0.3 dex, accounting for uncertainties in stellar population and extinction. Figure~\ref{fig1} shows the distribution of redshift versus stellar mass for our sample of 33 merging/interacting galaxies with strong tidal tails.

\begin{figure}[t]
  \centering
 \includegraphics[width=0.48\textwidth,bb=87 369 545 699]{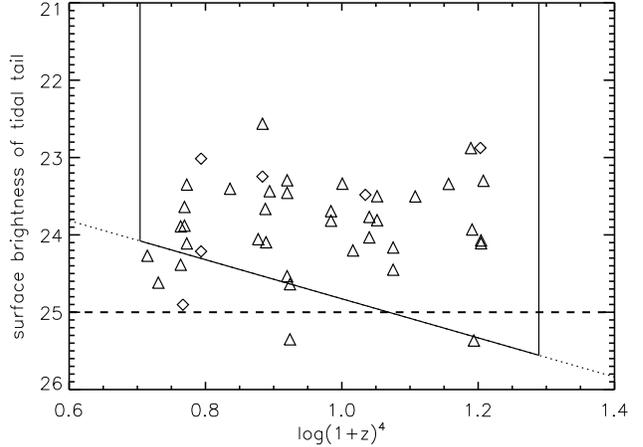}
 \caption{The tidal tail surface brightness as a function of redshift for our sample of 33 interacting/merging galaxies. The surface brightness is measured from ACS/$z$-band image. Of the 33 galaxies, 11 have two tidal tails. The dotted line shows the cosmic dimming with redshift for a target with surface brightness of 22.3\,mag\,arcsec$^2$ at $z=0$. The dashed line represents the depth of GEMS images. The triangles represent data points from the GEMS and the diamonds represent data points from the GOODS. The solid lines show redshift bin at $0.5<z<1.1$. \label{fig2}}
\end{figure}

  In addition,  we also examined detection completeness for tidal tails. The IRAF task ``POLYPHOT'' is used to perform polygonal-aperture photometry with ACS $z$-band image and measure the observed $z$-band surface brightness of the tidal tails for the selected 33 merging/interacting galaxies. The aperture is chosen to enclose tail regions brighter than 2 times the sky background RMS, and contain no stellar clump. Figure~\ref{fig2} shows the distribution of $z$-band surface brightness for tidal tails of 33 merging/interacting galaxies. Note that part of the sample galaxies have two tidal tails. The dashed line shows the GEMS image depth (i.e., 2 times the sky RMS) of 25\,mag\,arcsec$^{-2}$. Considering that the cosmic dimming with redshift as $(1+z)^4$ for the surface brightness of a fixed target, a tidal tail with intrinsic surface brightness of 22.3\,mag\,arcsec$^{-2}$ is detectable out to $z=0.85$ (see Section 3.3).

\subsection{TDG Identification}

  We visually identified stellar clumps as TDG candidates in tidal debris of merging/interacting galaxies. Circular aperture photometry was performed to the clumps.  Stellar mass is estimated using the same method for their parent galaxies.  A candidate will be treated as a TDG rather than super star cluster (SSC), young massive cluster (YMC) or giant HII region if it satisfies: 1) effective radius $R_e>$0.5\,kpc; 2) the central surface brightness (within $R_e$) 1.2 times brighter than that of the associated tidal tails and 3) stellar mass $> 10^7$\,M$_\odot$.

\begin{figure}[t]
  \centering
 \includegraphics[width=0.48\textwidth,bb=93 375 543 695]{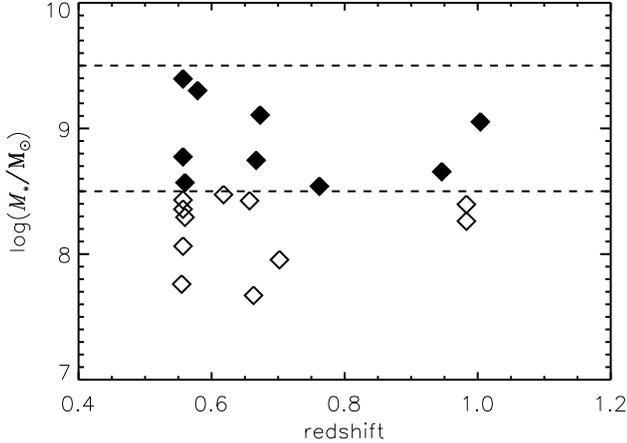}
 \caption{The stellar mass distribution of 20 TDGs candidates. The dash lines represent our stellar mass cut for massive TDG candidates. % while the dot line represents our redshift cut for our merging galaxy sample from 0.45 to 0.85. 
The filled diamond represent 9 massive TDG sample in our stellar mass cut.\label{fig3}}
\end{figure}

  There are 20 TDGs identified from the 33 sample galaxies. Figure~\ref{fig3} shows the stellar mass distribution of the 20 TDGs. Of them, 9 TDGs have stellar mass $3\times 10^8$\,M$_\odot<M_*<3\times 10^9$\,M$_\odot$. Table~1 lists the details for the 9 TDGs.

%No TDG candidate is found to be massive than $3\times10^8$\,M$_\odot$  at $z < 0.5$. 

\begin{figure}[t]
  \centering
 \includegraphics[width=0.48\textwidth,bb=70 370 545 698]{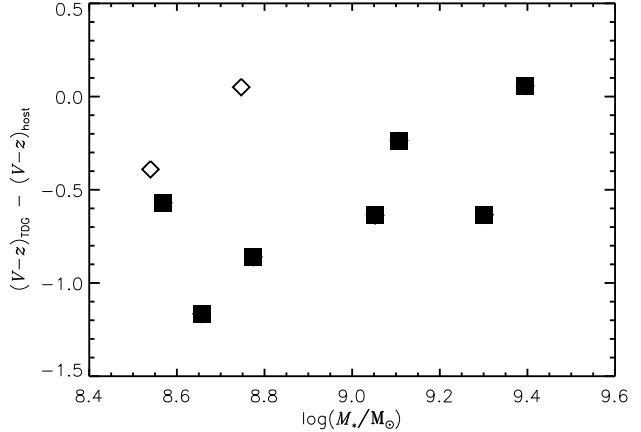}
 \caption{The difference of $V-z$ color between TDGs and their hosts versus stellar mass of the 9 massive TDGs. Seven TDGs from our subsample of 15 massive merging/interacting galaxies are shown with solid square.\label{fig4}}
\end{figure}

  Furthermore, we examined the $V-z$ color for the 9 massive TDGs. Figure~\ref{fig4} shows the $V-z$ color difference for the 9 massive TDGs from their host galaxies as a function of TDG stellar mass. A negative value of the deviation in $V-z$ means a bluer color of a TDG than its host galaxy. It is clear that TDGs are on average bluer than their host galaxies by about 0.5\,mag. This is likely suggestive of existence of a younger stellar population in TDGs if dust extinction plays a marginal role in shaping the observed color for both TDGs and their host galaxies. Although the uncertainties in dust extinction preclude any solid conclusion, the systematic offset in $V-z$ color suggests that the star formation activities are not identical for TDGs and their host galaxies.

  Here we identified TDGs according to their size, stellar mass and position. Contamination from background galaxies is still possible as warned by Weilbacher et al. (2003) when we selected stellar clumps within or at tip of tidal debris in merging sample. We calculated the number density of foreground and background galaxies with comparable z-band apparent magnitudes ($m_z$) to our TDG sample from the GOODS-S catalog. For $24~<~m_z~<~27$, we found that there are 0.039 galaxies per square arcsecond in GOODS-S. Tidal tails associating with our 58 merging galaxies sample totally cover 112 square arcseconds. This means that 2.8 out of 20 stellar clumps which visually embed in these tidal tails can be misidentified as TDGs, which cause an uncertainty of 14\% in the resulted production rate of long-lived TDG.

\subsection{TDG Production Rate} \label{yieldrate}

  Our goal is to measure TDG production rate per merger. A complete sample of parent merging/interacting galaxies is thus needed to estimate TDG production rate. The COMBO-17 catalog of galaxies with $R<24.1$ is complete for galaxies with stellar mass $\log (M^\ast/$\,M$_\odot)>10.5$ out to $z=1$.  Note that there are few massive galaxies in the underdense volume of the ECDFS at $z<0.5$ because of cosmic variance. We chose surface brightness cut of 22.3\,mag\,arcsec$^{-2}$ for tidal tail detection in order to enclose more sample galaxies. The detection is only complete up to $z<0.85$ for TDGs associated with tidal tails above this cut. By doing so, we finally selected a subsample of 15 merging galaxies with $0.5<z<1.1$, $\log (M^\ast/$\,M$_\odot)>10.5$ and $\Sigma _{\rm tail}<22.3$\,mag\,arcsec$^{-2}$. These selection criteria are shown in Figure~\ref{fig1} to \ref{fig2}. For this sample, the observed $z$-band corresponds to the rest-frame $B$ to $V$-band. The subsample contains 7 massive TDGs listed in Table~1. This results in about 0.47 massive TDGs per merger. Our sample is incomplete at $z>0.85$ in terms of the tidal tail surface brightness cut. We also estimated TDG production rate based on the subsample of $0.5<z<0.85$, leading to a production rate of 0.55 (5 massive TDGs among 9 mergers). This is consistent with the result based on the entire sample within uncertainties. We thus concluded that our estimate of TDG production rate is not significantly biased by the incompleteness due to the tidal tail surface brightness cut. Here we assumed that all wet mergers undergo a phase with bright tidal tails and wet mergers without clear tidal tails are in phases that the tidal tails fade away or do not appear yet (Mihos 1995). The TDG production rate estimated here is actually applicable for all mergers.

% To estimate the influence from detection limit at $0.85<z<1.1$, We re-calculated the TDG production rate from a smaller but complete merger sample in a narrower redshift bin ($0.5<z<0.85$). We found such complete subsample contain 9 mergers and 5 massive TDGs. This lead to a comparable TDG production rate (0.55 massive TDGs per merger) with larger uncertainty due to the smaller size of complete merger sample. We concluded that the detection limit of GEMS and a bright initial surface brightness cut of tidal tail did not lead to a serious influence on our calculation of TDG production rate from $0.5<z<1.1$ merger sample.}

%{\bf Moreover, note that our merging/interacting galaxies are specified to have strong tidal tails ($\Sigma _{\rm tail}<22.3$\,mag\,arcsec$^{-2}$) and this indeed leads a selection bias. Assuming that all wet merger (spirals-participated) will undergo phases of tidal tail from formation to fading away, however, we thought of our TDG production rate as a representative value for general TDG production rate over all kinds of mergers.} 
%which needs to be corrected when the specific production rate is converted to a general production rate of TDGs over all kinds of mergers. 

\begin{table*}[t]
\small
\caption{Photometric catalog of 9 massive TDGs.}
\begin{center}
\begin{tabular}{@{}cccccccccccc@{}}
\tableline
TDG&R.A.&DEC.&Host galaxy&redshift&$z_{850}$&$V_{606}-z_{850}$&$\log$ $(M_*/$M$_{\odot})$ & $\log$ $(M_*/$M$_{\odot})$ & size &\\   & $h~~ m~~ s$&$^\circ$ $'$ $''$& COMBO~~ID &  & mag & mag&TDG&Host galaxy & kpc & \\
\hline
1& 03 ~31 ~44.45 & -28 ~03 ~04.68 & 1984  & 0.762 & 25.86 & 0.68 & 8.54 & 10.05 & 1.55  \\
% 2& 03 ~31 ~24.95 & -28 ~02 ~41.09 & 2760  & 1.281 & 26.61 & 0.32 & 8.67 & 10.58 & 1.38  \\
2& 03 ~31 ~20.15 & -27 ~58 ~18.02 & 12222 & 1.004 & 25.00 & 0.64 & 9.05 & 11.09 & 1.44  \\
3& 03 ~33 ~19.08 & -27 ~56 ~57.76 & 15040 & 0.667 & 25.48 & 0.82 & 8.75 & 10.00 & 1.37  \\
4& 03 ~33 ~06.13 & -27 ~56 ~44.02 & 15599 & 0.56  & 24.81 & 0.41 & 8.57 & 10.85 & 1.84  \\
% 6& 03 ~32 ~01.78 & -27 ~52 ~41.71 & 23667 & 1.151 & 25.81 & 0.68 & 8.96 & 10.50 & 1.98  \\
% 7& 03 ~32 ~01.29 & -27 ~52 ~38.57 & 23667 & 1.151 & 25.71 & 0.56 & 8.91 & 10.50 & 2.22  \\
5& 03 ~31 ~37.90 & -27 ~50 ~14.34 & 28841 & 0.673 & 24.62 & 0.82 & 9.11 & 10.84 & 2.53  \\
% 9& 03 ~33 ~16.52 & -27 ~46 ~52.47 & 35611 & 1.256 & 26.19 & 0.24 & 8.70 & 10.52 & 1.39  \\
6& 03 ~31 ~58.81 & -27 ~44 ~58.88 & 39805 & 0.557 & 24.49 & 1.32 & 9.39 & 10.83 & 1.74  \\
7& 03 ~31 ~58.55 & -27 ~45 ~01.93 & 39805 & 0.557 & 24.29 & 0.41 & 8.77 & 10.83 & 3.48  \\
8& 03 ~32 ~11.27 & -27 ~42 ~40.31 & 45115 & 0.579 & 24.19 & 1.02 & 9.33 & 11.24 & 2.96  \\
9& 03 ~32 ~31.07 & -27 ~35 ~31.07 & 60582 & 0.946 & 25.25 & 0.06 & 8.66 & 10.69 & 1.19  \\

\tableline
\end{tabular}
\end{center}
%$\ast$: Massive clump ($10^9$\,M$_\odot$) lies at the base of tidal tail of COMBO 12222. This is unusual for TDGs at low redshift.
\end{table*}

\subsection{Contribution to the Local Dwarf population}

  We further estimated what fraction of dwarf galaxies in the local universe could be tidally originated. The total number of TDGs in a given co-moving volume at the present day depends on the number of major mergers per co-moving volume in the past, the production rate of TDGs per merger, and the average survival rate of a TDG over a given time.
  
  Jogee et al. (2009) presented the measurements of merger rate for galaxies with  $M^\ast > 2.5\times10^{10}$\,M$_\odot$ in four redshift bins from 0.24 to 0.8 in the ECDFS. The merger rate was estimated through $R_{\rm merger} = (n \times f_{\rm merger})/t_{\rm visual}$, where $n$ is the co-moving number density of field galaxies above a certain mass limit in a redshift bin; $f_{\rm merger}$ is fraction of mergers in the mass-limited galaxies; and $t_{\rm visual}$ is the timescale within which a merger exhibits visible key features for merger identification (0.5\,Gyr was adopted). %A time scale of 0.5\,Gyr was adopted by Jogee et al. (2009). Our nearly complete subsample includes 9 massive merging/interacting galaxies at $0.5<z<0.85$. Meanwhile there are at most 17 major mergers identified in the same redshift bin by Jogee et al. (2009) based on the same dataset as we used. Our massive major merger sample is smaller due to the higher stellar mass cut (0.1 dex) and further constraint on tidal tail brightness. As we discussed in section 2.2, the formation and fading away of tidal tail are two phases of dynamical evolution of a merger with comparable time scale. We thought that a similar TDG production rate can be found in mergers with faint tidal tails which are ignored in Elmegreen et al. (2007). So we need not do correction for  the specific production rate from our subsample because such value is equal to the general production rate of TDGs over all kinds of mergers.
In this work, we took the major merger rates over the redshift range $0.24<z<0.8$ from Table 1 of Jogee et al. (2009). Minor mergers are ignored because TDGs are thought to be formed only in major wet mergers (at least one disk-dominated galaxy involved). No correction was included for dry mergers because they are negligible compared to wet mergers in the ECDFS \citep{Bell06}. The major merger rate is 1.3$\times10^{-4}$\,per Gyr per Mpc$^3$ at $0.62<z<0.8$ and declines at decreasing redshift. By integrating the merger rate over the cosmic epoch from $z\sim 1.1$ to the present-day (8\,Gyrs), we obtained in total $6.8\times10^{-4}$ massive major mergers per Mpc$^3$. We cautioned that merger rates derived from different works show a large scatter mostly due to the limitation of methodologies and observational data. For instance,  the CAS (Concentration, Asymmetry, Clumpiness) morphological criterion \citep{Conselice03b} for selecting merging galaxies tends to misidentify irregular galaxies as mergers; close pair method relies on the merging timescale and correction for projection effect. Nevertheless, a global agreement on merger rates holds between recent measurements based on different methods and data (see Jogee et al.~2009 for reference there in; see also Robaina et al.~2010),  consistent with theoretical predictions from the Millennium Simulation \citep{Genel08}, Mare Nostrum simulation \citep{Dekel09},  semi-analytic models (Bower et al. 2006; Somerville et al. 2008), and cosmological SPH model (Maller et al. 2006) within large uncertainties. Considering the uncertainties in estimating merger rates, we realized that the estimated integrated number density of major mergers may vary by a factor 3.

Generally speaking, the mass ratio between TDGs and parent galaxies is $10^{-2}\sim10^{-3}$ or less. 
Because of the tidal friction with the parent galaxies, only massive TDGs can live for a long time. %The host galaxies should be more massive than serveral $10^{10}$\,M$_\odot$ if a TDG in these systems. 
Based on 96 N-body simulations of galaxy collisions with representative parameters (e.g., mass ratio and encounter geometry),  Bournaud \& Duc (2006) presented TDG survival rate as a function of time. We took their results and estimated the present-day TDG number density produced since $z=1.1$ by integrating galaxy merger rate, TDG production rate and survival rate over the examined cosmic epoch. We assumed that TDG production rate derived in this work is constant at all cosmic epochs. We adopted galaxy merger rates from Jogee et al. (2009). 
%\begin{equation}
%n_{\rm TDG} = \int_{0}^T (R_{\rm production})(N_{\rm merger})(R_{\rm survival})dt
%\end{equation}
%n_{\rm TDG} = 3.5\times10^{-5}(\frac{N_{\rm merger}}{6\times10^{-4}})(\frac{R_{\rm yield}}{0.23})(\frac{R_{\rm survival}}{0.25}),
%where $N_{\rm merger}$ is the integrated number of major mergers per unit volume; $R_{\rm yield}$ is the production rate of TDGs per merger; $R_{\rm survival}$ is the adopted survival rate of TDGs over 7\,Gyrs.
%{\bf where $R_{\rm production}$ is the production rate of TDGs per merger and we assumed here that it is uniform at any redshift; $N_{\rm merger}$ is the number of major mergers per unit volume at a given redshift bin; $R_{\rm survival}$ is the adopted survival rate of TDGs at a given redshift bin; `` $T$ '' is the up limit of the lookback time.} Combining TDG production rate from our estimate with TDG survival rate from Bournaud \& Duc (2006) and galaxy major merger rates from Jogee et al. (2009), 
Finally we calculated the accumulated numbers per unit volume for TDGs with mass $10^{8.5-9.5}$\,M$_\odot$ produced in the past 8\,Gyrs (i.e., since $z=1.1$) to be $8.2\times10^{-5}$ per Mpc$^3$. This number density increases by a factor of 2 to account for TDGs produced before $z=1.1$ by assuming a survival rate of about 18\% (to the present day) and a merger rate of $2\times10^{-4}$\,Gyr$^{-1}$\,Mpc$^{-3}$ for $1.1<z<4$ \citep[e.g. Extrapolation from][]{Conselice03, Conselice06}. 
%this value is variable according to the assumed survival timescale; $n_{total}$ is the number density of dwarf galaxy with two origins in local universe. 
% This lead to totally $6.8\times10^{-5}$ massive TDGs has survived during last 7Gyr and evolution into isolated system in per $Mpc^3$ local universe if the production rate of long lived massive TDG is about 0.1 per merger (section 3). 
From the local galaxy stellar mass function \citep{Bell03} we derived the number density of local dwarf galaxies in stellar mass range from $3\times10^8$\,M$_\odot$ to $3\times10^9$\,M$_\odot$ as $7\times 10^{-3}$\,Mpc$^{-3}$. Putting all together, we simply derived that TDGs contribute $<\sim 3\%$ to the local dwarf galaxy population. Note that the uncertainty in production rate of TDG is dominated by Poisson noise due to a small number of sample galaxies and contributed by the TDG misidentification due to the contamination from foreground/background galaxies. The survival rate for TDGs may differ from the adopted value by a factor 2 (Hunsberger et al. 1996). Accounting for uncertainties in the integrated number density of major mergers, the production rate and survival rate of TDG, we suggested that less than 10\% dwarf galaxies in the local universe are of tidal origin.

%$N_{\rm merger}$, $R_{\rm production}$  and $R_{\rm survival}$, 

%\emph {The uncertainty in $F_{\rm TDG}$ is considered as following. According to the Poisson noise in number count of our complete subsample of merger and TDG misidentification due to the contamination from foreground/background galaxies, $R_{\rm yield}$ has an uncertainty of 72\%. Uncertainty in the integrated number of massive major mergers per local unit volume from last 7\,Gyr can be 75\% according to merger rate calculation in Jogee et al. (2009). }

\section{SED Modeling for Two TDGs}

\begin{figure*}[t]
  \centering
 \includegraphics[width=0.65\textwidth,bb=8 43 496 596]{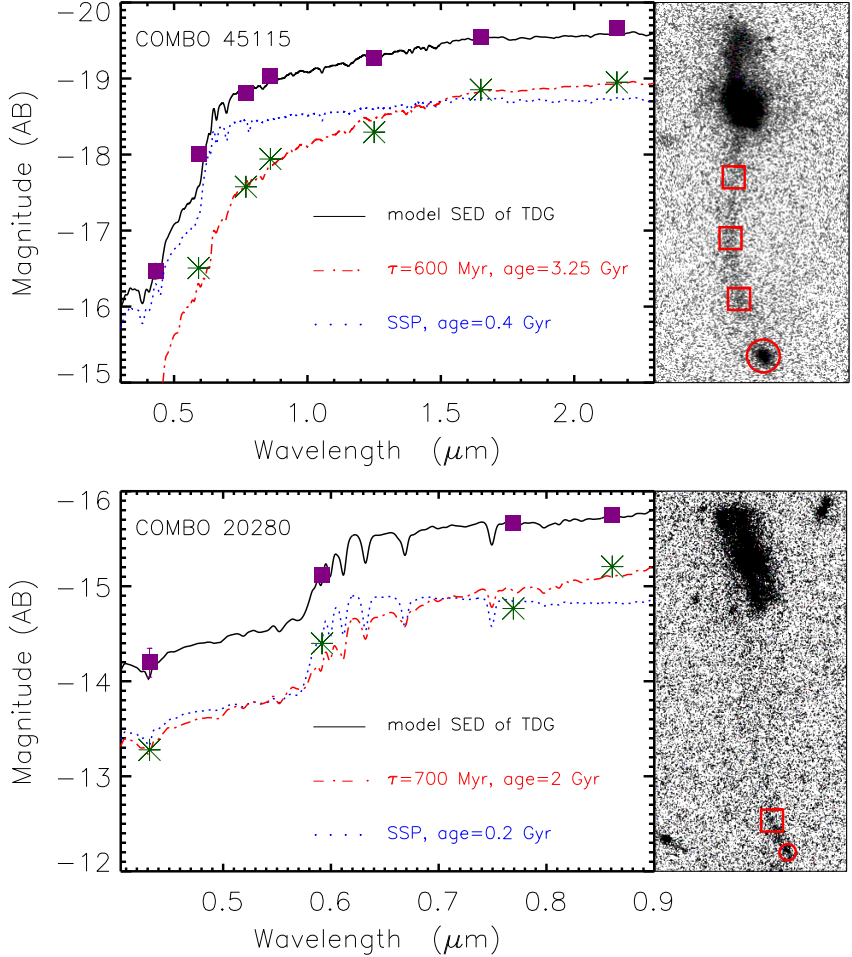}
 \caption{The observed SEDs of TDG (squares) and tidal tail (stars) for TDG 45115 (upper) and TDG 20280 (bottom).  The right panels show morphologies of TDGs and tidal tails with photometry apertures marked. The best-fit stellar population models are also shown. See text in Section~4 for details of the modeling. \label{fig5}}
\end{figure*}

Two TDGs identified surrounding our sample galaxies 20280 and 45115 (COMBO-17 ID) are located in the GOODS-S field. The existing multi-band HST/ACS observations from the GOODS survey allow us to build up broadband spectral energy distributions (SEDs). Moreover, the object 45115 has near-infrared (NIR) $J,H$ and $Ks$ images observed with the Infrared Spectrometer And Array Camera (ISAAC) on board the VLT \citep{Retzlaff10}. The optical-to-NIR SEDs provide observational constraints on SFH for the two objects and hints for the difference between tidally-originated and cosmologically-originated dwarf galaxies.
Morphologies of the two galaxies are shown in Figure~\ref{fig5}. Galaxy 45115 shows double nuclei, a long bright tail toward the South and a warped tail toward the North. Galaxy 20280 exhibits a long tail. 
Both of the two galaxies contain a TDG at the end of the tidal tail.
We used the available data to derive the broadband SEDs for the two systems.

\subsection{Aperture-Matched Broadband SEDs}

We constructed optical SEDs from four-band HST/ACS observations.  
Because the image quality and resolution of the $B, V, i$, and $z$-band images are very similar, we adopted an aperture of radius 0.$''$6/0.$''$24 to integrate all signals of two given TDGs (in 45115/20280 respectively) in all ACS images.
Similarly, we derive NIR SEDs from  ISAAC $J,H$ and $Ks$-band images for TDG in galaxy 45115. 
An aperture of radius 1$''$ was adopted. We then derived aperture-matched color $z-Ks$ to link optical and NIR SEDs together. The aperture matching was done following the method developed by Wuyts et al. (2008). Firstly we combined 5$\times$5 pixel block in ACS $z$-band image into one pixel for the different pixel size between ACS and ISAAC image. Empirical PSFs were constructed using 62 bright, isolated and unsaturated stars selected from COMBO-17 and EIS catalogs \citep{Groenewegen02} for ACS $z$-band ( after rebin) and ISAAC $Ks$-band images. The empirical PSFs were fitted with a Moffat profile in order to filter noise out (particularly in the wing regions). The best-fit Moffat profile is defined by ($\beta = 3.31, FWHM=0.''18$) and ($\beta = 3.23, FWHM=0.''54$) for the rebined ACS $z$-band and ISAAC $Ks$-band PSF, respectively. A convolution kernel was created with the smoothed empirical $z$ and $Ks$-band PSFs \citep[see][for more details]{Zhang10}. Rebined ACS $z$-band image was  then convolved with the kernel to match the $Ks$-band image. Finally the aperture-matched $z-Ks$ color was derived from the PSF-matched images using the same aperture for the $Ks$-band photometry. Figure~\ref{fig5} shows the aperture-matched SEDs (squares) for TDG 45115 (upper) and TDG 20280 (bottom). Meanwhile, the aperture-matched SEDs are also derived for tidal tails. 
The photometry for the tidal tail of galaxy 45115 was done over three separated square regions along the tail in order to improve signal-to-noise ratio. Sky background was determined from parallel regions close to the apertures. 
The square apertures are marked on galaxy images. % in the middle panels of Figure~\ref{fig5}. 
For comparison, the SEDs of tidal tails are plotted along with SEDs of corresponding TDGs. We can see from Figure~\ref{fig5} that in both cases TDG is typically 0.5\,mag bluer than tidal tail, consistent with the results shown in Figure~\ref{fig4} for all TDGs.

\subsection{Stellar Population Modeling}

  A typical TDG usually contains two stellar components. One component is the pre-existing stellar population from the TDG's parent galaxy. The other is a newly formed stellar population triggered by the interaction/merger. 
 The stellar population from the parent galaxy had been formed before the collision happened, while the gas, if sufficient, can form a young stellar component during the collision \citep[e.g.][]{Weisz08, Sabbi08}. The associated tidal tail is composed of the same old stellar population from the parent galaxy.
In order to quantify the contribution of young stellar component to the total in a TDG, we analyzed the broadband SEDs presented in Figure~\ref{fig5} using two component (i.e., old+young) models.
The data used here lack the resolution to probe SFH for the old stellar population. Instead, we took the observed SED of corresponding tidal tail as representative of the old stellar population. A single stellar population (SSP) with solar metallicity and Chabrier IMF from the evolutionary population synthesis model BC03 is used to represent the young stellar component. No extinction is counted. 
The stellar mass of the old component, the age and stellar mass of the SSP are free parameters in fitting the observed SEDs. The best-fit model was chosen using the least squares method. 
For demonstration, we fitted tidal tail's SED with galaxy models of e-folding formation history. 
Figure~\ref{fig5} shows the fitting results for the TDG in galaxy 45115 (upper) and the TDG in 20280 (bottom). For TDG 45115, the best-fit model for the tidal tail is a stellar population formed through e-folding ($\tau=0.6$\,Gyr) with age 3.25\,Gyr, and an SSP with age 400\,Myr is favored for the young stellar component.
The stellar masses for the young and old components in TDG 45115 are $8\times 10^8$\,M$_\odot$ and $1.2\times 10^9$\,M$_\odot$, respectively. 
The total stellar mass derived from SED fitting is similar to the estimate from color-luminosity method. The newly-formed stars in TDG 45115 contribute roughly 40\%  to the total stellar mass.
For TDG 20280, the observed SED favors a model consisting of a young stellar component of age 200\,Myr and mass $1.5\times 10^7$\,M$_\odot$ and an old stellar component of age 2\,Gyr and mass $4.3\times10^7$\,M$_\odot$. The young stars contribute about 26\% to the total stellar mass. 

The SED analysis for the two TDGs strongly suggests that the color offset shown in Figure~\ref{fig4} is caused by  a stellar population newly formed during galaxy collision. 
We emphasized that our results from the stellar population analysis of SEDs are consistent with these obtained from the color-magnitude analysis of resolved stellar population for TDG Holmberg IX in M81 group \citep{Sabbi08} and for TDG Arp 105S \citep{Duc97}.

\section{Discussion and Conclusion}

  We combined HST/ACS imaging data from GOODS-S and GEMS surveys with photometric redshifts from COMBO-17 survey to carry out a morphological analysis for a sample of 33 
%massive (stellar mass $>3\times 10^{10}$\,M$_\odot$ ) 
merging/interacting galaxies with strong tidal tails at $0.5<z<1.1$ in the ECDFS. The imaging observations adopted here allow us to detect a tidal tail of surface brightness down to 25\,mag\,arcsec$^2$ (2 times sky background RMS). Accounting for the cosmic dimming effect, tidal tails with surface brightness $<22.3$\,mag\,arcsec$^{-2}$ are detectable out to $z=0.85$. The sample galaxies were carefully examined and 20 stellar clumps were identified as TDG candidates. Every candidate is connected with its host galaxy through an apparent tidal tail. We finally obtained 9 TDGs satisfying: (1) effective radius $R_{\rm e}>0.5$\,kpc; (2) central surface brightness (within $R_{\rm e}$) 1.2 times higher than that of the associated tidal tail; (3) stellar mass $\sim 10^{8.5-9.5}$\,M$_\odot$.
  In order to estimate the production rate of long-lived TDGs produced in merging/interacting of massive galaxies, we introduced further constraints on stellar masses for host galaxies and TDGs. A subsample of 15 merging/interacting galaxies with stellar mass $>3\times 10^{10}$\,M$_\odot$ were selected in the redshift range of $0.5<z<1.1$. Accordingly, the subsample contains 7 TDGs with stellar mass $3\times 10^8$\,M$_\odot < M_\ast < 3\times 10^9$\,M$_\odot$ identified from tidal tails of intrinsic surface brightness brighter than 22.3\,mag\,arcsec$^{-2}$. We thus estimated that the production rate of massive TDGs as about 47\% per merger.
% of specific tidal tail in length and surface brightness. By comparing to the merger rate measurements from Jogee et al. (2009), we corrected the production rate by a factor of 0.7 here.

  Taking the major merger rates since $z=1.1$ from Jogee et al. (2009) and survival rate from Bournaud \& Duc (2006), we estimated the contribution of TDGs to local dwarf galaxy population. Our estimates (see Section~3 for more details) show that the total number of TDGs with mass $10^{8.5-9.5}$\,M$_\odot$ produced in the past 12\,Gyrs (i.e., since $z=4$) and survived to the present day is $1.5\times10^{-4}$ per Mpc$^3$. The number density of local dwarf galaxies with $3\times10^8$\,M$_\odot <M_\ast <3\times10^9$\,M$_\odot$ was derived from the stellar mass function given by Bell et al. (2003). Uncertainties in the integrated number of major mergers per unit volume, the production rate of TDGs per merger and the survival rate are considered. Putting all together, the contribution of TDGs to the present-day dwarf galaxy population in the field is less than 10\%.

It is worth noting that the fraction of dwarf galaxies to be tidal originated could be higher in dense environments (e.g., groups and clusters) or around early-type galaxies where more frequent interactions or mergers happened in the past. 
For instance, Hickson compact groups \citep[HCGs,][]{Hickson82} are likely to host a high fraction ($\sim 56$\%) of TDGs \citep{Hunsberger96}.
As densest concentration of galaxies, HCGs are comparable to the central regions of rich clusters in galaxy density ($10^4\sim10^6$\,Mpc$^{-3}$) but characterized with a lower velocity dispersion leading to more collision and interaction events\citep{Hickson92}. 
%
%Hickson compact groups \citep[HCGs,][]{Hickson82} which consist of 451 galaxies within 100 groups can provide a satisfactory place to study such problem. They are among the densest concentrations of galaxies known, comparable to the centers of rich clusters ($10^4\sim10^6$/Mpc$^3$). However, they also have relatively low velocity dispersions ($\sigma\sim250$ km/s, smaller than what is seen in rich clusters). The combination of high densities and low velocity dispersions should lead to more collision and interaction events between the galaxies in these groups \citep{Hickson92}.
%
Detailed studies have shown much evidence for abundant merger events in HCGs \citep[e.g.,][]{Rubin91,Zepf91,Zepf93,Moles94}.
Tidal features such as bridges and tails were found in nearly half of HCG member galaxies (Mendes de Oliveira \& Hickson 1994). The much steeper faint end of luminosity function also suggests a significant fraction of dwarf galaxies in HCGs tend to be of tidal origin \citep[][]{Hunsberger98}.

We emphasized that the TDG formation processes at high-z ($\sim 2-4$) might be quite different from low-z ($<1$). Spiral galaxies at high-z are generally gas rich and clumpy, and with star formation rate of order 100\,M$_\odot$\,yr$^{-1}$ (Genzel et al. 2006; F\"{o}rster Schreiber et al. 2006; Genzel et al. 2008; Stark et al. 2008). Star-forming clumps in high-z spirals are found to be much more massive than typical star-forming clouds in local spirals \citep{Elmegreen07b,Elmegreen09}. Such clumps could form through gravitational instabilities (Elmegreen et al. 2004; Elmegreen \& Elmegreen 2005; Shapiro et al. 2008; Bournaud et al. 2008; Daddi et al. 2010). The LCDM models are actually able to reproduce such clumpy disk galaxies at high-z (Ceverino et al. 2010). Interestingly, Bournaud et al. (2011) showed that merger involving disk galaxies with a number of giant clumps induce interaction and scatter between these giant clumps and some of them can be expulsed far from the merger and stay aside as long-lived ($>\sim 2$\,Gyrs) massive TDGs. 
% Interaction between these clumps could make them dissolve, lose angular momentum, migrate inward, finally evolve into a central bulge and an exponential disk like typical spiral galaxies today \citep{Bournaud07}.
%
% Merger/interaction between these clumpy galaxies would present special dynamics which differ from mergers at low redshift. Such process has been firstly studied in simulation work by Bournaud et al. (2010). It shows that some of these clumps will be expulsed relatively far and become new, long-lived ($\sim 2$\,Gyr ) TDGs of $10^8$\,M$_\odot\sim10^9$\,M$_\odot$ after the merger. They claimed that such process can be a new and efficient TDG forming mechanism.
%
If such dynamical process often occurs in mergers of gas-rich disk galaxies at high-z, one would expect more long-lived TDGs produced in each high-z merger by throwing some of clumps away, compared to a typical gas-rich merger at $z<1$. This gives rise to an additional mechanism for TDG formation, implying a potentially higher fraction of local dwarf galaxies to be tidally formed. 
The compact environment coupled with such efficient TDG forming mechanism could result in a ``TDG-boom'' in the early epoch of the universe \citep[also see][]{Kroupa10}. 
The production rate and survival rate for these TDG-like objects at high-z have not yet been systematically explored. More efforts are needed to investigate TDG formation at $z>1$.

  We built up the aperture-matched broadband SEDs for two TDGs associated with parent galaxies COMBO 45115 and 20280 in the GOODS-S. We used two-component (young+old) stellar SED models to fit the observed SEDs.  Spectral templates with solar metallicity and Chabrier IMF from BC03 were adopted. The best-fit results suggest a young component of age 400\,Myr and stellar mass 40\% of the total for TDG 45115 and a young component of age 200\,Myr and stellar mass 26\% of the total for TDG 20280. Combined with a systematic bluer color shown in Figure~\ref{fig4}, our results indicate that TDGs often form a stellar population in a relatively short timescale (usually several 10$^8$\,Myrs). We argue that merging/interacting processes are responsible for such episodic star formation activities in TDGs.

\acknowledgments
We thank the anonymous referee for the valuable comments to improve the paper.
Z.\ Z.\ Wen is grateful to Ran Wang, Zhiyu Zhang for helpful discussion and specially to Hongxin Zhang for kindly providing the image convolution code. This work was supported by the National Natural Science Foundation of China (10773030,10833006) and the National Basic Research Program of China (2007CB815404).

%figure1
\newpage

\newpage

\end{document}